# Semiconductor Detector Developments for High Energy Space Astronomy

**Aline Meuris**[a]

[a] *Commissariat à l'énergie atomique et aux énergies alternatives,
CEA Saclay, 91191 Gif-sur-Yvette, France
E-mail*: `aline.meuris@cea.fr`

ABSTRACT: The rise of high energy astrophysics and solar physics in the 20th century is linked to the development of space telescopes; since the 1960s they have given access to the X-ray and gamma-ray sky, revealing the most violent phenomena in the Universe. Research and developments in imaging concepts and sensing materials haven't stopped since yet to improve the sensitivity of the X-ray and gamma-ray observatories. The paper proposes an overview of instrument realizations and focuses on the innovative detection techniques and technologies for applications from 0.1 keV to 10 MeV energy range. Solid-state detectors are prominent solutions for space instrumentation because of their excellent imaging and spectroscopic capabilities with limited volume and power resources. Various detection concepts based on semiconductors (Compton camera, Cd(Zn)Te pixel hybrids, DePFET active pixel sensors) are under design or fabrication for the near-future missions like Astro-H, BepiColombo, Solar Orbiter. New technologies on sensing materials, front-end electronics, interconnect processes are under study for the next generation of instruments to push back our knowledge of star and galaxy formation and evolution.

# Contents

# 1. Introduction

High energy astronomy refers to the detection of radiation coming from the sky with energy greater than 1 keV ($10^{-16}$ J). High energy gamma-rays (E > 10 GeV) can be detected on ground when they interact high up in the atmosphere and generate an air shower of secondary particles that produce Cherenkov light. Telescope arrays like HESS [1], MAGIC [2] and in the future CTA [3] have been designed for that purpose. But X-rays and low energy gamma-rays are mainly absorbed by our atmosphere (by 50% at 30 km altitude for 1 MeV photons) and require the development of space telescopes. From the 1950's, gamma-ray sources were predicted, based on particle and nuclear physics theories and first discoveries in radioastronomy; but the difficulty to embed sufficiently large detection areas in balloons and sounding rockets used to limit observations. Experimental X-ray and gamma-ray astronomy really started at the end of 1960's, with the American SAS (Small Astronomy Satellite) and OSO (Orbiting Solar Observatory) series satellites. X-ray emissions from Scorpius X-1 and Centaurus X-3 were interpreted as neutrons stars accreting matter from an orbiting star: the concept of X-ray binaries was born. From 1980 to 1989, the Solar Maximum Mission provided the first hard X-ray and gamma-ray spectra of solar flares and gave quantitative diagnostics of energetic particles interacting at the Sun (Bremsstrahlung continuum emission by electrons, nuclear line emission by neutron capture and positron annihilation). Major breakthrough for gamma-ray astrophysics came later, at the beginning of the 1990's, with the Compton Gamma-ray Observatory (CGRO) and the Sigma telescope onboard the Granat satellite. Various stellar objects at the end of their evolution (neutron stars, black holes, supernovae) were detected and observed; Batse onboard CGRO provided the very first catalog of gamma-ray bursts. Active galactic nuclei —whose gamma-ray emission comes from the gravitational energy produced during matter accretion around a supermassive black hole— were identified in few percents of the galaxies.

This paper reviews high-energy photon detectors and associated space instruments developed since then and presents mature technologies for the future observatories. Section 2 introduces the detector types and design rules for space applications and highlights the importance of solid-state detectors. Section 3 reviews instrumentation for X-ray astronomy, based on grazing incidence mirrors and silicon detectors. Section 4 reviews instrumentation for gamma-ray astronomy, for which indirect imaging techniques and various detection concepts had to be invented to optimize the sensitivity of the telescopes; state of the art of germanium and Cd(Zn)Te detectors are presented. Performance status and development perspectives are given in conclusion.

## 2. Detectors in space for high energy astronomy

### 2.1 Scientific needs and technical solutions

The astronomers need both imaging capability to identify the X-ray and gamma-ray sources in the sky and spectroscopy to understand the complex physics of these objects: spectral signature gives access to temperature and acceleration processes in the active regions (accretion disks, jets). Polarimetry should give precious information about the geometries of magnetic field and emission regions in compact objects and binary systems. For solar physicists, photon spectrum gives access to the primary electron spectrum revealing particle acceleration and energy transfer processes; imaging solar flares is also a way to understand the structures of electrical and magnetic fields at the surface of the Sun.

High-energy detectors are based on single photon detection, by measuring the charges created by the ionization or the temperature elevation (micro calorimeter). We can classify the photodetectors in 4 categories:

- **Gaseous ionization detectors**: used in few space experiments like Integral/JEM-X, they require a large volume for good efficiency and they are sensitive to voltage breakdown by cosmic proton interaction. They could be of interest for X-ray polarimetry [4].
- **Scintillation detectors**: first detectors used in space, coupled with photomultipliers (NaI onboard OSO-7 in 1972). Inorganic scintillation crystals stay efficient solutions for energy measurement in gamma-ray experiments above 10 MeV because they can be implemented in large quantities and volumes (e.g. recently in AGILE [5] and FERMI [6]). Plastic scintillators are also widely used as active shielding in high-energy experiments for background particle detection and rejection.

- **Semiconductor detectors**: Si, Ge, CdTe, CdZnTe detectors have progressively replaced scintillators in space experiments because of their better intrinsic spectral performance; they show now high technological maturity to propose compact solutions for both high resolution imaging and spectroscopy from the visible range to the soft gamma-ray range, with versatile readout, dimensions and granularity.
- **X-ray calorimeter**: shorty operated in space (helium failure after calibration onboard Suzaku in 2005), they would provide ultimate energy resolution, ~50 times better than solid-state detectors to resolve features of the iron line fluorescence features at 6 keV in compact objects. R&D programs are on-going based on superconducting transition-edge-sensors or metal insulator semiconductor coupled with high-Z absorbers [7]. They require cryogenic temperatures (<100 mK), which makes the realization and the readout of matrices extremely challenging.

**2.2 Development of a space instrument based on semiconductor detectors**

The choice of a detector type and geometry depends on performance requirements (energy range, detection efficiency, detection area, spatial resolution, spectral resolution, time resolution) and allocated resources, which are limited in a spacecraft (10 to 100 W electrical power, 10 to 100 kg mass budget, 0.1 to 1 Mbit per day telemetry downlink for a typical instrument). In addition to semiconductor detector properties, space environment has to be taken into account to design all subsystems of an instrument.

**The sensor part**

Solar protons and electrons trapped in earth radiation belts induce background particles that make the detectors blind to sky observation. To optimize the observation time, astronomy missions prefer low-earth orbits with small incidence angle, or elliptical orbits or L2 Lagrange point. Background particles ($10^5$ to $10^9$ 10 MeV-equivalent protons/cm$^2$ over a space mission duration) contribute to the ageing of the semiconductor detectors in space by creating lattice displacement damage. The lattice defects may act as traps or recombination centers for charge carriers. First consequence is an increase of leakage current that limits the spectral resolution; second consequence is a decrease of the mobility × lifetime product of the charge carriers that affects charge collection efficiency. To limit these effects, the sensors have to be carefully shielded; cooling down and increasing the electrical field in the sensor are also recommended. For intrinsic material like germanium, in-flight annealing at ~100°C is an efficient way to suppress most lattice defects and to recover the performance.

**The front-end electronics**

Recent designs of front-end electronics are Application Specified Integrated Circuits (ASICs) with mixed electronics for most of them: analog front-end channels with charge sensitive preamplifier, first filtering stage and digital electronics for readout control, signal sampling. Electronic noise performance is sensitive to the detector leakage current and the input capacitance and front-end electronics has to be optimized to the sensor properties; some detector laboratories develop full custom ASIC for that purpose [8,9,10]. The ASIC shall withstand the hostile radiation environment in space: background particles (protons essentially) are responsible for single events, which can create failures from bit flip (upset) to complete destruction (latch-up, gate ruptures, burnouts…). The total ionizing dose (equivalent to hundreds of grays for standard space missions) provokes degradation of the transistor properties (threshold voltages), which generally impacts the performance of the analog part.

**The hybridization**

To optimize performance, the sensor is closely connected to the front-end electronics, with various manufacturing and interconnects processes. The hybridization techniques can be inherited from the semiconductor manufacturing technologies (wire bonding, bump bonding, flip chip) but the resulting hybrid detector shall be qualified to the space environment (vibrations and shocks during launch, temperature variations in orbit).

**The shielding**

Last design specificity of instruments for high-energy astronomy concerns the shielding of the detector that directly impacts the sensitivity performance of the telescope. Heavy shielding stop X-rays and gamma rays out of the field of view whereas active and light shielding detect charged particles and reject events seen in coincidence in the detector plane.

## 3. X-ray telescopes: focusing imaging and silicon based devices

### 3.1 Imaging techniques

X-rays are not reflected by classical optical mirrors. Hans Wolter proposed in 1952 three configurations to focus X-rays using grazing incidence mirrors [11]. A Wolter type-I telescope consists of a hyperboid mirror and a paraboloid mirror to focus X-rays after 2 consecutive reflections (see Fig. 1). Due to the very low limit incidence angle for reflection, the collection area is restricted to a thin ring. To optimize the effective area and the volume of the mirror, mirrors shells of different diameters are mounted in a coaxial and confocal way.

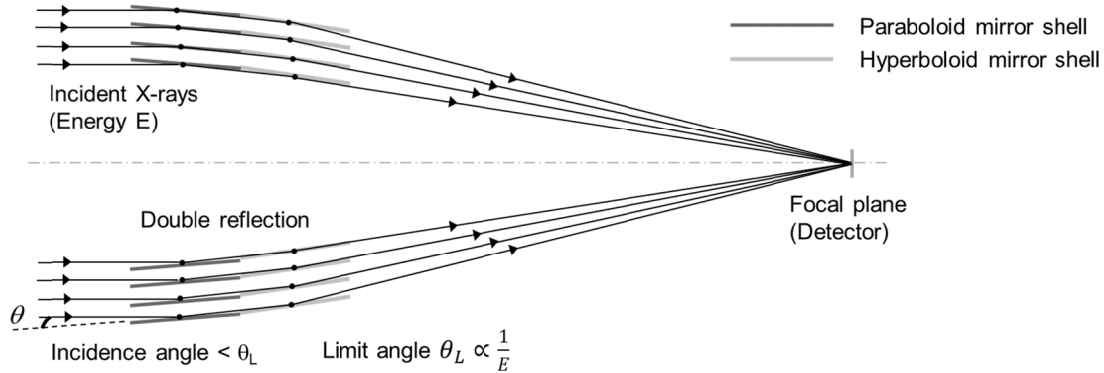

**Figure 1**: Principle of the X-ray focusing imaging with mirrors in Wolter-I configuration.

The first grazing mirror was launched in 1979 on-board the Einstein Observatory to focus X-rays until 4 keV. Since then, a number of X-ray or gamma-ray missions built at the end of the 20th century have flown with Wolter-I configuration mirrors or an approximation of it: ROSAT (1990), SAX [12] (1996), Chandra (1999), XMM-Newton [13] (1999), Swift (2004). Typical performances of a telescope unit are 5 arcsec HEW (half energy width) angular resolution at 1 keV, 12 arcmin field of view, 1000 cm$^2$ effective area at 1 keV (500 cm$^2$ at 6 keV) and $5.10^{-8}$ ph.s$^{-1}$.cm$^{-2}$.keV$^{-1}$ continuum sensitivity at 1 keV (3σ point source detection, 100 ks exposure). By increasing the focal length above 10 m and adding a multilayer coating on the mirrors to increase the reflection index, it is now possible to focus hard X-rays like in NuSTAR launched in 2012 [14] (120 cm$^2$ at 50 keV). Research on mirror technology is on-going to limit costs and mass of the optical system: microchannel plate detectors, radially packed in Wolter-I configuration, will fly on-board BepiColombo for Mercury exploration [15] in 2016 and silicon pore optics are in study at ESA to reach few arcsec angular resolution at 6 keV for a future large X-ray observatory [16].

### 3.2 From charge coupled devices to active pixel sensors

#### 3.2.1 MOS CCD and pnCCD for X-ray imaging spectroscopy

Charge-coupled devices (CCD) appeared in the 1970's and were rapidly adopted by astronomers for optical imaging in ground-based telescopes. Conventional devices consist of metal-oxide-semiconductor (MOS) structures on top of a p-type silicon wafer (see Fig. 2a). Signal charge is stored in potential wells underneath the oxide and then transferred to a collecting electrode by a suitable change of the gate potentials. Two extra features are now integrated in all CCD to limit the charge transfer inefficiency (CTI); a thin n-doped layer between silicon and oxide (n-buried channel) prevents the trapping of electrons by the surface crystal defects; the p-type strips implanted perpendicular to the gate electrodes (channel stops) prevents charge spreading towards the adjacent rows during transfer. High dynamic range, high quantum efficiency and low noise make CCD suitable for X-ray imaging spectroscopy; matrices ranging from 100k to 1M pixels were developed by e2V and MIT's Lincoln Laboratory to equip focal planes of X-ray telescopes on-board ASCA [17] (1993), Chandra, XMM-Newton [18], Swift and Suzaku (2005). Energy resolution of 75 eV FWHM at 523 eV (O line) and 130 eV FWHM at 6 keV (Fe line) can typically be obtained cooling the sensors below −50°C. Limitations of MOS CCD come from the thin depletion layer (30 to 65 µm) that implies low quantum efficiency above few keV. The pnCCD is an alternative concept based on the double diode structure with sideward depletion, alike the silicon drift detectors [19] (see Fig. 2b). The device is thus fully depleted and can be back illuminated to improve quantum efficiency over a large energy range (0.2- 10 keV). Produced by the semiconductor laboratory of the Max-Planck-Institut, they were used in XMM-Newton [20] and foreseen for the eROSITA telescope [21] and the X-ray telescope of the SVOM gamma-ray burst (GRB) mission [22].

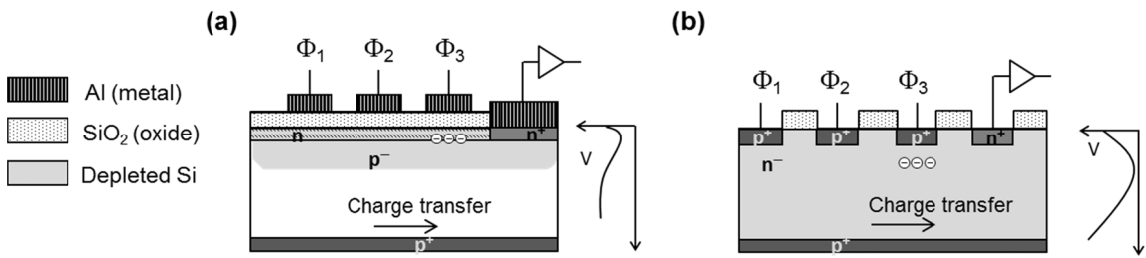

**Figure 2**: (a) Structure of a buried-channel MOS-CCD. Charges accumulated few microns below the oxide interface before being transferred row wise using three potential phases. The depletion area is limited to 30 to 65 µm. b) Structure of a pnCCD. The device is fully depleted (250 to 500 µm) and can be illuminated from the backside to limit absorption by the dead layers of the pixel structures.

### 3.2.2 Active pixel sensors in development

Taking advantage of the expansion of the semiconductor industry at the end of the 20th century, new detector developments aim at combining high resistive silicon material for X-ray detection with doped silicon for front-end electronics in hybrid detectors. The charge collection efficiency is no longer affected by radiation damage in space since there is no transfer in the bulk to read out the signal. The first stage of amplification in the pixel guarantees low input capacitance and thus, excellent energy resolution.

One relevant example of active pixel sensors (APS) for space applications is the focal plane of the MIXS-T for BepiColombo made of 64 × 64 macropixels of 300 μm by side. A macropixel is the association of a square-shaped silicon drift detector (SDD) with a depleted p-channel field effect transistor (DePFET). The drift ring structure creates a potential well that makes the electron cloud (created in the pixel bulk by photon interaction) drift quickly towards the pixel center, in a potential minimum underneath the FET (see Fig. 3). The FET channel conductivity is a function of the amount of charges in the so-called internal gate. The charge is stored until applying a positive voltage on the clear voltage to remove it from the internal gate. Such APS can be associated with various readout schemes: rolling shutter mode, on-demand readout (to select regions of interest) or repetitive readout (to reduce measurement noise) since the readout is not destructive.

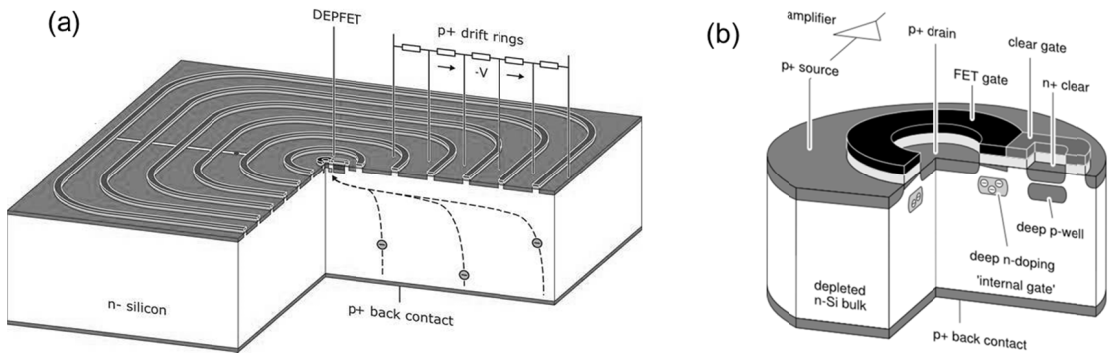

**Figure 3**: Structure of a macropixel of the APS realized for BepiColombo MIXS-T, combining a SDD structure for fast collection over 300 μm pixel size and a DePFET structure for first amplification and non-destructive readout (by courtesy of pnSensor).

Macropixel structures are innovative solutions to adapt the pixel size to the point spread function of the focusing optics. For fine imaging, the pixel structure can be limited to the DePFET itself (50 to 100 μm pixel size); the concept has been proposed for the wide field imager of the Advanced Telescope for High Energy Astrophysics [23] (ATHENA+, formerly IXO, selected as the second L-class mission of the ESA Cosmic Vision program). For fast timing and high resolution spectroscopy, the SDD structure, already implemented in industrial systems for various spectroscopic applications, could be used for black hole and transient X-ray sources observations, like foreseen for the Large Observatory For X-ray Timing (LOFT), candidate for a ESA M-class mission [24].

# 4. Gamma-ray telescopes: modulation imaging techniques and high-Z detectors

## 4.1 Imaging techniques

Since focusing techniques are not applicable at high energy, imaging concepts based on multiplex collimation schemes were developed for hard X-rays and gamma-rays, consisting of 'encoding' the photon signal temporally or spatially on the detectors.

### 4.1.1 Coded-mask technique for all sky monitor

For high energy astronomy, the coded-aperture systems are the best solution to provide large field of view for sky survey and good sensitivity to detect faint sources; they acts like multiple pinhole camera, coding the information of the incoming photons in the form of overlapping images on the detector (see Fig. 4). Among the possible patterns of opaque and transparent elements, the uniformly redundant arrays proposed by Fenimore and Cannon in 1978 [25] allow the source position reconstruction (imaging by matrix deconvolution) and minimizes the influence of background. Such masks were used for Granat/SIGMA (1989), Integral/IBIS [26] (2001) and Swift/BAT [27] (2002). Typical best performance of a coded-mask telescope is $3.10^{-6}$ $ph.s^{-1}.cm^{-2}.keV^{-1}$ sensitivity at 100 keV (3σ, 100 ks), 1 sr field of view and 12 arcmin angular resolution.

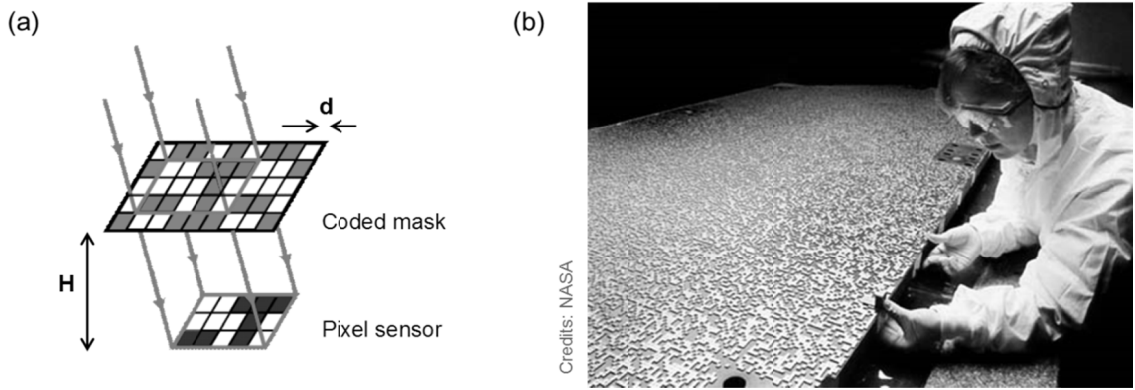

**Figure 4**: (a) Principle of coded-aperture telescope for gamma-ray sky imaging. The image projected on the detector plane depends on the point position of the source at the infinite. The angular resolution is defined by *d/H*. This technique requires detectors with the same spatial resolution of the mask pattern, typically 2 mm. (b) Picture of the coded-mask of Swift/BAT made of ~54000 lead tiles of 5 mm × 5 mm × 1 mm with 50% opening area. The mask has an area of 2.7 $m^2$ yielding a half-coded field of view of 1.4 sr.

### 4.1.2 Rotating collimators for solar flare imaging

For solar physics, arcsec angular resolution is needed to resolve the solar flares and would require μm spatial resolution detector with a coded aperture system of few meters long. A widely used imaging technique is based on the spatial Fourier transform [28]. The telescope consists of several collimators, made of two grids with parallel linear slits, separated by a distance *D* which is large compared with the slit width *s*. Each collimator allows the measurement of a Fourier complex visibility, in a similar way as interferometry with ground based radiotelescopes; the angular frequency is given by the *D/s* ratio, independent of the detector spatial resolution. By using grid pairs with a variety of slit spacings and orientations, the source distribution is sampled in the (u,v) plan of spatial frequencies at a variety of points. The variation of orientations can be realized by rotating the collimators with respect to the source, as done for the RHESSI mission [29] (2002) and several balloon experiments; non-rotating collimators were used for the Hard X-ray Telescope of Yohkoh (1991) and will be used for the Spectrometer Telescope Imaging X-rays in Solar Orbiter [30] (2017). Typical performance is 2 to 8 arcsec resolution up to 100 keV and full sun field of view (1°).

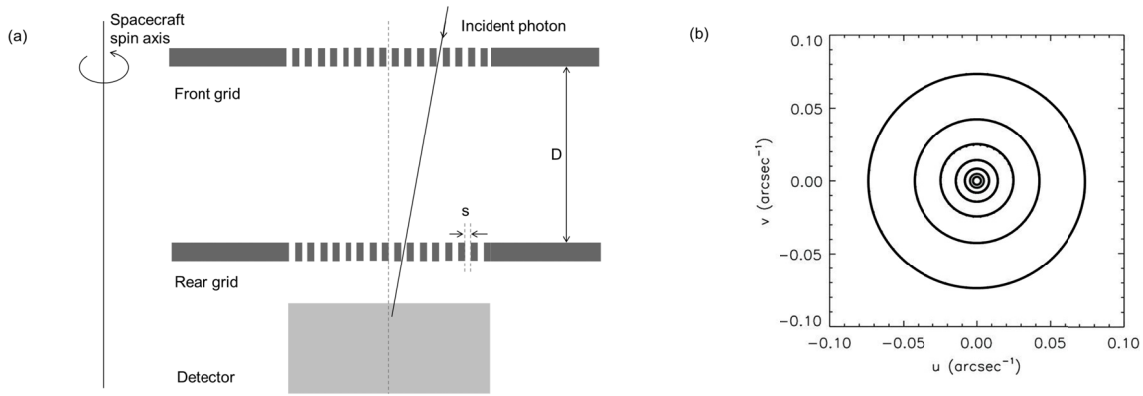

**Figure 5**: Principle of X-ray imaging with rotating modulation collimators. (a) One collimator is made of a pair of grids and measures one complex visibility in the (u,v) plan of spatial frequencies; the amplitude of the visibility is given by *s/D*. The rotation of the spacecraft modulates the signal in time to access to different phases in the (u,v) plan. (b) Resulting sampling of the image distribution in the (u,v) plan by the RHESSI collimators #3 to #9. Image reconstitution (equivalent to an inverse Fourier transform) can be done by the same techniques as in back projection or other methods from radio interferometry.

### 4.1.3 Compton telescope

When the photoelectric effect is no longer predominant (E > 100 keV in Ge, E > 200 keV in CdTe), Compton imaging technique is an alternative solution to optimize detection efficiency. The principle is to use two layers of sensors to detect high-energy photons, one to Compton scatter the primary photon, one to absorb the scattered photon. The Compton kinematics gives access to the incident angle (see Fig. 6a) and the incident energy can be obtained if the two layers are position sensitive spectrometers; moreover, if the photon statistics is enough, the distribution of the azimuthal angle can give access to the angle and the fraction of light polarization (see Fig. 6b). First Compton telescopes were realized with scintillators like Comptel on-board CGRO in 1991 [31]; the JAXA Institute of Space and Astronautical Science has designed semiconductor Compton cameras with silicon detectors as scatter layers and CdTe detectors as absorbers. The Soft Gamma-Ray Detector of Astro-H to be flown in 2014 is the stacking of 32 Si layers and 8 CdTe layers of CdTe and surrounding them with 2 CdTe layers [32]. Typical best performance of such imaging systems is 3° angular resolution (limited by Doppler broadening), $4.10^{-7}$ $ph.s^{-1}.cm^{-2}.keV^{-1}$ sensitivity over 100-500 keV (3σ, 100 ks) sensitivity (thanks to good background rejection by the Compton kinematics), 1% minimum detectable polarization for bright sources (Crab nebula, 100 ks).

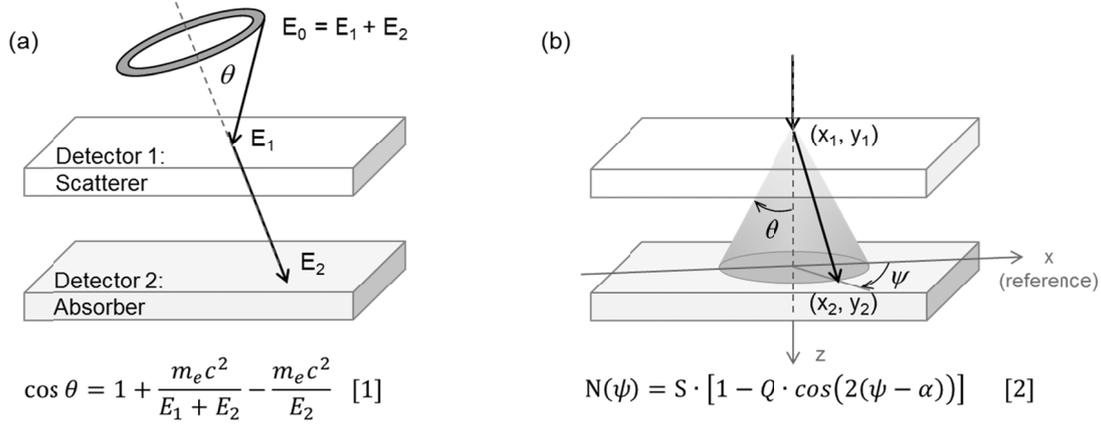

**Figure 6**: (a) Principle of a Compton camera. The incident angle is given by the formula [1]. (b) If the source is linearly polarized with an angle $\alpha$ with respect to the *x* axis, the photon distribution along the azimuthal angle $\psi$ is modulated according to the formula [2].

### 4.2 Germanium and Cd(Zn)Te Detectors: from monolithic crystals to high spatial resolution devices

#### 4.2.1 Planar detectors

High purity germanium detectors (GeD) have been used in gamma-ray instruments that did not require fine spatial resolution, in particular the Spectrometer of Integral [33] and the RHESSI spectrometer. They can achieve high-resolution spectroscopy over 4 decades of energy (5 keV - 15 MeV). For the hard X-ray detection planes of the coded-aperture telescopes in Integral and Swift, position sensitive sensors were 2 mm-thick Cd(Zn)Te crystals for an energy range from 15 to 150 keV. The Integral Soft Gamma-Ray Imager (ISGRI) is an array of 2621 cm$^2$ realized by assembling detection units called polycells; a polycell consists of 16 mono-pixel 4 mm × 4 mm CdTe crystals and four analog front-end ASICs mounted on a ceramic board. The design and detector geometry were similar for the 5012 cm$^2$ CdZnTe detection plane of Swift. Development of CdTe detectors with low leakage current (Schottky diodes) and improvement in front-end electronics performance make now possible to realize the same kind of detection plane with a low detection threshold of 3 to 4 keV, as foreseen for SVOM/ECLAIRs telescope to reveal GRBs at cosmological distances.

#### 4.2.2 Double-sided strip detectors

Processes of electrode deposition and patterning have improved for both Ge and Cd(Zn)Te detectors and position sensitive sensors are now widely used. The finest spatial resolution can be achieved with strip contacts; processing both sides of a planar detector with orthogonal strips give a 2D positioning. This can be done on high purity germanium by implanting boron on one side and diffusing lithium on the other side to create pn junctions. Typical pitch of 2 mm can be obtained conventionally. The Lawrence Berkeley Laboratory has developed a process of high resistivity amorphous semiconductor coating followed by a metal patterned electrode deposition on top of it to reach pitch of 0.5 mm [34]. Such devices are foreseen to equip of a Compton telescope and a Fourier-transform telescope in balloon experiments [35].

Pitch down to of 60 μm have been obtained with metal electrode lithography on Cd(Zn)Te detectors. Double-sided strip detectors have the advantage of having less readout channels (2N) than a pixel array (N$^2$), possibly an easier connection scheme (at the sensor periphery) and hence, the possibility of layer stacking. For the Hard X-ray Imager (HXI) of Astro-H integrating both double-sided strip silicon and CdTe detectors [36], the strips are connected a ceramic board by gold-stud bonding and then connected to a multi-channel front-end ASIC using through holes and wire bonding (see Fig. 7a).

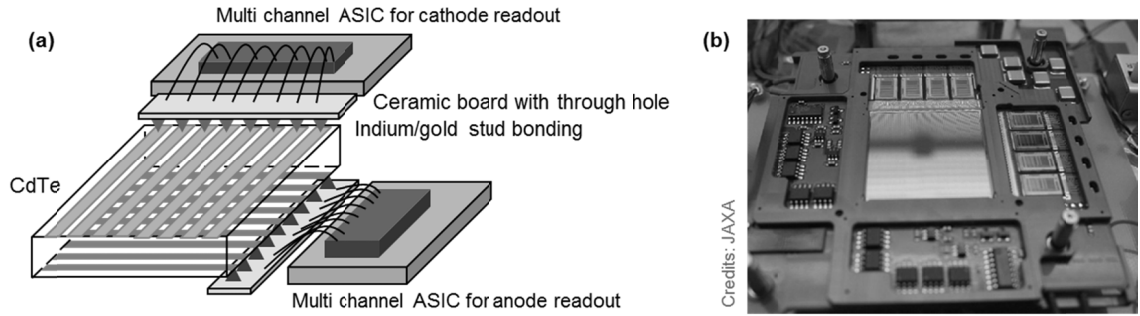

**Figure 7**: (a) Hybridization principle of a double sided strip CdTe detectors and a multi-channel analog front-end ASIC for Astro-H/HXI. (b) Engineering model of a double sided silicon strip detectors with 4 VA32TA6 ASIC (including ADC) on each side for Astro-H/HXI.

### 4.2.3 Pixel Cd(Zn)Te hard X-ray cameras

The major drawbacks of DSD are interaction position ambiguity in case of multiple events and large input capacitance of strips that prevents from achieving ultimate spectral performance (not better than 1 keV FWHM at 60 keV). For hard X-ray focusing optics, new concepts of Cd(Zn)Te hybrid detectors with ~600 µm pitch have been developed. One design proposed by CEA-Saclay consists in putting the front-end ASICs perpendicular to the detector array in an electronic body and connecting the pixels using flex leads and polymer bump bonding. The resulting 4-side buttable hybrid called Caliste was developed in the prospect of a large focal plane (> 20 cm$^2$) [37] (see Fig. 8a). For the NuSTAR hard X-ray telescope launched in 2012, the California Institute of Technology designed a 2D ASIC with front-end channels in 498 µm pixel sizes. The CZT pixel sensor is connected to 2 ASICs with conductive epoxy and gold stud using a flip chip bonding technique. The ASICs are glued and connected by wire bonding to an interface board holding the analog to digital converters; the result hybrid is 3-side buttable (see Fig. 8b). Energy resolution better than 0.8 keV FWHM at 60 keV and 1.5 keV low-energy threshold can be obtained over the array in both designs.

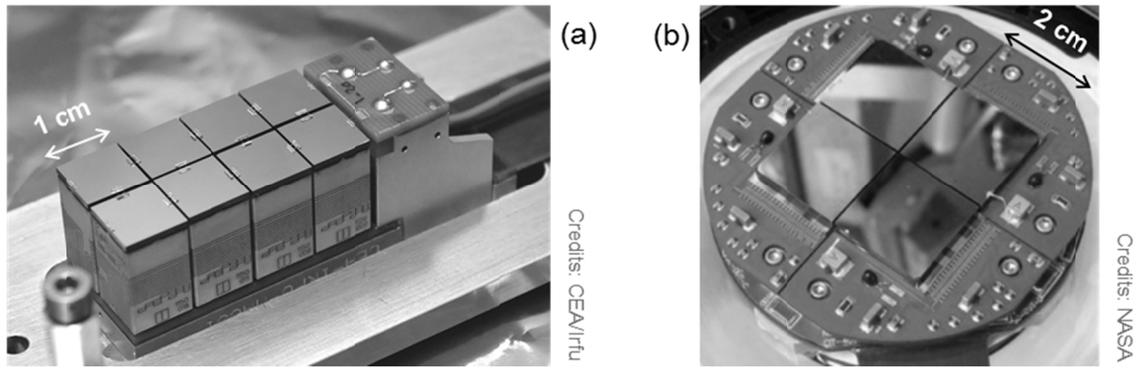

**Figure 8**: (a) Assembly of 8 Caliste-HD CdTe hybrid detectors (8 cm$^2$, 2048 pixels, 625 μm pitch). Same hybridization technology (CEA/3D Plus) is foreseen for spectrometers of Solar Orbiter/STIX and Interhelioprobe/Sorento. (b) NuSTAR focal plane made of 4 CZT hybrid detectors (16 cm$^2$, 4096 pixels, 605 μm pitch).

## 5. Conclusions

Based on imaging techniques invented at the end of the 20$^{th}$ century, the detector concepts have continuously multiplied based on semiconductor technologies. For experiments from soft X-rays (0.2 keV) to soft gamma-rays (0.5 MeV), silicon and Cd(Zn)Te materials offer the possibility of realizing various compact cameras by stacking detection layers or combining sensor with front-end electronics in monolithic detectors (APS). Most instruments in implementation phase take advantage of special care of the entrance window and new designs of low noise rad-hard front-end electronics to reach very low-energy threshold (0.1 keV with Si, 1 keV with Cd(Zn)Te). Energy resolution obtained is closed to the intrinsic limit of the Fano statistics (2% at 6 keV with Si, 1% at 60 keV with Cd(Zn)Te). Thanks to mature detector technologies and after addressing sensitivity, angular and spectral resolution, next generation telescopes could improve timing performance to study transient X-ray source and solar bright events and polarimetry capability to understand black hole physics. New technological challenges consist then in building megapixel detection plane to come closer to what is already available for the optical telescopes; that requires the production of uniform wafer-scale silicon detectors, the bonding of Cd(Zn)Te pixel array with 2D ASICs, and probably new integration concepts using 3D technologies to stack functional layers (sensor, ASIC, ADC, controller interface) into smart hybrids – all of it taking into account the issue of power heat dissipation and space mechanical and thermal environment.

For gamma-ray experiments (0.5 – 20 MeV), germanium material offers now the possibility of combining thick detectors and fine pitch resolution. Next adopted space missions don't plan to use them but balloon experiments are in preparation to prove the interest of these detectors for Compton telescopes and gamma-ray imaging polarimeters; meanwhile research and development on curved crystals is on-going to build Laue diffraction lens [38]. The arrival of an instrument with a gamma-ray focusing imaging system would bring breakthrough for nuclear astronomy.

**Acknowledgments**

The author thanks Dr. Olivier Limousin and Dr. Philippe Laurent for their comments on this review.

## References

[1] J.A. Hinton, *The status of the H.E.S.S. project*, New Astron.Rev. **48** (2004) 331-337.

[2] D. Ferenc, *The MAGIC gamma-ray observatory*, NIM A **553** (2005) 274-281.

[3] J. Hilton, S. Sarkar, D. Torres and J. Knapp, *Seeing the High-Energy Universe with the Cherenkov Telescope Array – The Science Explored by the CTA*, Astrop. Phys. **43** (2013) 1-356.

[4] R. Bellazini, *A sealed Gas Pixel Detector for X-ray astronomy*, NIM A **579** (2007) 853-858.

[5] M. Tavani et al., *Science with AGILE*, AGILE document AP-27 (2003) http://agile.mi.iasf.cnr.it/

[6] W. B. Atwood, et al., *The Large Area Telescope on the Fermi Gamma-Ray Space Telescope Mission*, ApJ **697** (2009) 1071.

[7] C. Enss, *Cryogenic Particle Detection*, Topics in Applied Physics **99**, Springer-Verlag (2005).

[8] O. Gevin et al., *Imaging X-ray detector front-end with dynamic range: IDeF-X HD*, NIM A **695** (2012) 415-419.

[9] K. Oonuki et al., *Development of Uniform CdTe Pixel Detectors Based on Caltech ASIC*, Proc. of SPIE **5501** (2004) 218-228 [arXiv:astro-ph/0410040].

[10] M. Porro et al., *Performance of ASTEROID: a 64 channel ASIC for source follower readout of DEPFET matrices for X-ray astronomy*, Proc. IEEE NSS Conf. Rec. (2008) 1830-1835.

[11] H. Wolter, *Spiegelsysteme streifenden Einfalls als abbildende Optiken für Röntgenstrahlen*, Annalen der Physik **10** (1952) 94-114.

[12] L Scarsi, *SAX overview*, A&A Suppl. Ser. **97** (1993) 371-383.

[13] D. de Chambure et al., *XMM's X-ray Telescopes*, ESA Bulletin 100 (1999) 30-42.

[14] F. A. Harrison et al., *The Nuclear Spectroscopic Telescope Array (NuSTAR)*, ApJ (2013) [arXiv:1301.7307].

[15] G.W. Fraser et al., *The Mercury imaging X-ray spectrometer (MIXS) on BepiColombo*, Planetary and Space Science **58** (2010) 79-95.

[16] M. Bavdaz et al., *Silicon pore optics development and status*, Proc. of SPIE. **8443** (2012) 844329.

[17] B. E. Burke, *An abuttable CCD imager for visible and X-ray focal plane arrays*, IEEE Trans. Electron Devices **38** (1991) 1069-1076.

[18] M. J. L. Turner et al., The European Photon Imaging Camera on XMM-Newton: the MOS cameras, A&A **365** (2001) L27-L35.

[19] G. Lutz, *Semiconductor Radiation Detectors*, Springer-Verlag Berlin Heidelberg (1999) chapter 6.

[20] L. Strüder et al.., *The European Photon Imaging Camera on XMM-Newton: the pn-CCD cameras*, A&A **365** (2001) L18-L26.

[21] P. Predehl et al, *eROSITA*, Proc. of SPIE **6266** (2006) 62660P.

[22] O. Godet et al., *The Chinese-French SVOM Mission: studying the brightest astronomical explosions*, Proc. of SPIE **8443** (2012) 84431O.

[23] A. Rau, *The Hot and Energetic Universe – the Wide Field Imager (WFI) for Athena+*, http://athena2.irap.omp.eu/IMG/pdf/SP_WFI_APH.pdf

[24] M. Feroci et al., *LOFT – a Large Observatory For x-ray Timing*, *Proc. of SPIE* **7732** (2010).

[25] E. E. Fenimore, and Cannon T. M., *Coded Aperture Imaging with Uniformly Redondant Arrays*, *Applied Optics* **17** (1978) 337-347.

[26] P. Ubertini et al., *IBIS: The Imager on-board INTEGRAL*, *A&A* **411** (2003) L131-L139.

[27] S. D. Barthelmy et al., The *Burst Alert Telescope (BAT) on the Swift MIDEX Mission*, *Space Sci. Rev.* **120** (2005) 143-164.

[28] T. A. Prince et al., *Gamma-Ray and Hard X-Ray Imaging of Solar Flares*, *Solar Physics* **118** (1998) 269-290.

[29] R. P. Lin et al., *The Reuven Ramaty High-Energy Solar Spectroscopic Imager (RHESSI)*, *Solar Physics* **210** (2002) 3-32.

[30] A. O. Benz et al., *The Spectrometer Telescope for Imaging X-rays (STIX) on board the Solar Orbiter mission*, *Proc. of SPIE* **8843** (2012) 8843131.

[31] V. Schönfelder et al., *Instrument Description and Performance of the Imaging Gamma-Ray Telescope COMPTEL aboard NASA's Compton Gamma Ray Observatory*, *ApJ Suppl.*, **86** (1993) 657.

[32] S. Watanabe et al., *Soft-Gamma-ray Detector for the ASTRO-H mission*, *Proc. of SPIE* **8443** (2012) 844326.

[33] G. Vedrenne et al., *SPI: the spectrometer aboard INTEGRAL*, *A&A* **411** (2003) L63–L70.

[34] M. Amman, P. N. Luke and S. Boggs, *Amorphous-semiconductor-contact germanium-based detectors for gamma-ray imaging and spectroscopy*, *NIM A* **579**, *(2007) 886-890*.

[35] A. Y. Shih et al., *The Gamma-Ray Imager/Polarimeter for Solar flares (GRIPS)*, *Proc. of SPIE* **8443** (2012) 84434H.

[36] M. Kokubun et al., *The Hard X-ray Imager (HXI) for the ASTRO-H mission*, *Proc. of SPIE* **8443** (2012) 844325.

[37] O. Limousin et al., *MACSI: Modular Assembly of Caliste spectroscopic imagers*, Radiation Detectors and their applications, *CRC Press* (accepted).

[38] F. Frontera and P. Von Ballmoos, *Laue Gamma-Ray Lenses for Space Astrophysics: Status and Prospects*, *X-Ray Optics and Instrumentation* **2010** (2010) 215375.